\documentclass[reprint,amsmath,amssymb,aps,pra]{revtex4-1}

\usepackage{mathtools}

\usepackage{graphicx}
\usepackage{dcolumn}
\usepackage{bm}
\usepackage{subcaption} 

\usepackage{hyperref}
\usepackage{url}

\usepackage[usenames,dvipsnames]{xcolor}
\colorlet{BLUE}{blue}
\colorlet{RED}{red}

\newcommand{\Mpl}{M_{\rm Pl}\,}

\begin{document}

\title{Improved bounds on ultralight scalar dark matter in the radio-frequency range}





\author{Oleg Tretiak}
\affiliation{Johannes Gutenberg-Universit{\"a}t Mainz, 55128 Mainz, Germany}
 \affiliation{Helmholtz-Institut, GSI Helmholtzzentrum f{\"u}r Schwerionenforschung, 55128 Mainz, Germany}
 
 \author{Xue Zhang}
\email{xuezhang@uni-mainz.de}
\affiliation{Johannes Gutenberg-Universit{\"a}t Mainz, 55128 Mainz, Germany}
 \affiliation{Helmholtz-Institut, GSI Helmholtzzentrum f{\"u}r Schwerionenforschung, 55128 Mainz, Germany}

\author{Nataniel L. Figueroa}
\affiliation{Johannes Gutenberg-Universit{\"a}t Mainz, 55128 Mainz, Germany}
 \affiliation{Helmholtz-Institut, GSI Helmholtzzentrum f{\"u}r Schwerionenforschung, 55128 Mainz, Germany}

\author{Dionysios Antypas}
\affiliation{Johannes Gutenberg-Universit{\"a}t Mainz, 55128 Mainz, Germany}
 \affiliation{Helmholtz-Institut, GSI Helmholtzzentrum f{\"u}r Schwerionenforschung, 55128 Mainz, Germany}

\author{Andrea Brogna}
\affiliation{Johannes Gutenberg-Universit{\"a}t Mainz, 55128 Mainz, Germany}

\author{Abhishek Banerjee}
\affiliation{Department of Particle Physics and Astrophysics, Weizmann Institute of Science, Rehovot, Israel 7610001}

\author{Gilad Perez}
\affiliation{Department of Particle Physics and Astrophysics, Weizmann Institute of Science, Rehovot, Israel 7610001}

\author{Dmitry Budker}
\affiliation{Johannes Gutenberg-Universit{\"a}t Mainz, 55128 Mainz, Germany}
 \affiliation{Helmholtz-Institut, GSI Helmholtzzentrum f{\"u}r Schwerionenforschung, 55128 Mainz, Germany}
\affiliation{Department of Physics, University of California, Berkeley, California 94720, USA}

\date{\today}

\begin{abstract}
 We present a search for fundamental constant oscillations in the range \mbox{20 kHz$-$100 MHz}, that may arise within models for  ultralight dark matter (UDM). Using two independent, significantly upgraded optical-spectroscopy apparatus,  we achieve up to $\times$1000 greater sensitivity in the search relative to previous work. We report no observation of UDM and thus constrain respective couplings to electrons and photons within the investigated UDM particle mass range \mbox{$8\cdot 10^{-11}-4\cdot 10^{-7}$ eV}. The constraints significantly exceed previously set bounds, and as we show, may surpass in future experiments those provided by equivalence-principle experiments in a specific case regarding the combination of UDM couplings probed by the latter.
\end{abstract}

\maketitle

\author{X. Zhang}
\affiliation{Johannes Gutenberg-Universit{\"a}t Mainz, Helmholtz-Institut Mainz, Mainz 55128, Germany}

\author{N. L. Figueroa}
\affiliation{Johannes Gutenberg-Universit{\"a}t Mainz, Helmholtz-Institut Mainz, Mainz 55128, Germany}

\author{D. Budker}
\affiliation{Johannes Gutenberg-Universit{\"a}t Mainz, Helmholtz-Institut Mainz, Mainz 55128, Germany}

\author{D. Antypas}
\affiliation{Johannes Gutenberg-Universit{\"a}t Mainz, Helmholtz-Institut Mainz, Mainz 55128, Germany}



\maketitle


\emph{Introduction --} One of the important quests of modern physics is understanding the nature of dark matter. Within a broad class of scenarios, dark matter is made of bosonic fields that are associated with light particles such as axions or axion-like particles, which are  classified according to their spin, interaction types with standard model (SM) matter and resulting observables \cite{PRESKILL1983127, ABBOTT1983133, DINE1983137,GrahamPRD2016}. They may have mass $m_{\rm{\phi}}$ in a broad range $10^{-22}-10$\,eV 
and form a classical oscillating field $\phi(t)\approx\phi_{0}\sin(2\pi f_{\phi}t)$, with the oscillation frequency being close to the Compton frequency of the particle $f_{\rm \phi}=m_{\rm{\phi}}/2\pi$ \footnote{We use natural units, where $\hbar=c=1$.}.
In cases where this ultralight dark matter (UDM) field $\phi$ has scalar coupling to SM matter, the interaction is expected to appear as an apparent oscillation in the fundamental constants (FC) occurring at the frequency $f_{\phi}$. It may also give rise to Equivalence-Principle-(EP)-violating acceleration \cite{GrahamPRD2016,HeesPRD2018}. Such scalar couplings are present within string/dilatonic theories \cite{ArvanitakiPRD2015}, and within beyond-SM  extensions introduced to explain the hierarchy problem \cite{GrahamPRL2015} that were further developed to accommodate the presence of UDM \cite{FlackeJHEP2017,BanerjeePRD2019, BanerjeeJHEP2020}.

Searches for effects of light scalar fields involve analysis of astrophysical data from the early universe \cite{StadnikFlambaumPRL2015, SibiryakovJHEP2020}, fifth-force experiments to probe EP-violation \cite{,SmithPRD1999, SchlammingerPRL2008,TouboulPRL2017,BergePRL2018} or apparent FC oscillations. The latter give rise to oscillations of specific atomic parameters that can be sensitively probed. For instance, the energy of atomic levels and thus the frequency of electronic transitions is approximately proportional to the Rydberg constant $R_{\infty}=(1/2)m_{\rm{e}}\alpha^2$, where $m_{\rm{e}}$ is the electron mass and $\alpha$ is the fine-structure constant. In addition, the length of solid bodies, which is proportional to the Bohr radius $\alpha_{\rm{B}} =(\alpha \, m_{\rm{e}})^{-1}$ depends on the same constants. 
Atomic and optical techniques are sensitive means to look for oscillations in $\alpha$ and $m_{\rm{e}}$ \footnote{ other methods involve use of mechanical \cite{ArvanitakiPRL2016, ManleyPRL2020} or acoustic resonators \cite{CampbellPRL2020}.}, for example,  by probing the frequencies of atomic transitions \cite{ArvanitakiPRD2015,SafronovaADP2019} in atomic clocks \cite{VanTilburgPRL2015, HeesPRL2016, WiczloSciAdv2018, SafronovaPRL2018}, and comparing these to the resonance frequency of optical cavities \cite{AharonyPRD2021, KennedyPRL2020}, via laser interferometry \cite{StadnikPRL2015,StadnikPRA2016}, comparison of two cavities \cite{GeraciPRL2019},  gravitational-wave detectors \cite{ArvanitakiPRD2018,GrotePhysRevResearch2019, VermeulenArxiv2021} and other methods \cite{SavallePRL2020, OswaldArxiv2021}. 

%

A method involving optical spectroscopy of an atomic ensemble to probe oscillations of $\alpha$ and $m_{\rm{e}}$ in the radio-frequency (rf) range 20\,kHz-100\,MHz \mbox{($8\cdot 10^{-11} < m_{\rm{\phi}}< 4\cdot 10^{-7}$ eV)} was introduced in Ref.\,\cite{AntypasPRL2019}.  In this  range, searches for EP-violating fifth forces have been more sensitive in exploring the scalar field. FC oscillations may be greatly enhanced, however, if there exist UDM halos that are gravitationally bound to the Earth  \cite{BanerjeeComPhys2020} or the Sun  \cite{BanerjeeComPhys2020, AndersonArxiv2020}. Such halos may result in an enhanced local DM density, and correspondingly, to enhancement of FC oscillations. 
In such cases, the observability of the effects of the scalar UDM field may be greater in the case of  FC-oscillation experiments compared to EP-violation ones. This is because an EP-violating fifth-force involves virtual exchange of the scalar particle, that is independent of UDM. 
There is another reason why direct UDM searches and EP tests can be considered complementary to each other: in part of the parameter space of UDM-SM couplings probed by EP tests, their sensitivity is reduced compared to direct searches \cite{OswaldArxiv2021}, as we discuss below.

Here we present an improved search for scalar UDM within the same mass range \mbox{($8\cdot 10^{-11}-4\cdot 10^{-7}$ eV)} as that explored in \cite{AntypasPRL2019}. Through the use of improved apparatus and techniques, we achieve  a substantially greater sensitivity in probing fast FC oscillations, and obtain constraints on the couplings of the scalar field that are improved by up to $\times 10^3$  with respect to \cite{AntypasPRL2019}. The sensitivity also significantly exceeds that of the recently reported results from the co-located optical interferometers (the Fermilab Holometer) \cite{aiello2021constraints} that cover part of the parameter space addressed by our experiments.

\emph{Experimental principle --} The idea to probe FC oscillations is to compare the frequency of an atomic transition $f_{\rm{at}}$ to the  frequency $f_{\rm{L}}$ of a laser field exciting it, and look for relative variations $\delta f=f_{\rm{at}}-f_{\rm{L}}$ \cite{AntypasPRL2019, AntypasQST2021}. With $f_{\rm{L}}$ tuned to excite the transition, $f_{\rm{L}}\approx f_{\rm{at}}$, and such variations occur because $f_{\rm{at}}$ and $f_{\rm{L}}$ have different dependence on the FC.
The dependence of the frequency $f_{\rm{i}}$ on a  constant $g$ can be quantified through the coefficient $Q^{\rm{i}}_{\rm{g}}=d\,\rm{ln}\,\mathit{f}_{\rm{i}}/ \,\mathit{d} \, ln\,\mathit{g}$ \cite{KozlovADP2019}. With this, one may write for the relative variation: \mbox{$(\delta f/f)_{\rm{g}}=(Q^{\rm{at}}_{g}-Q^{\rm{L}}_g)(\delta g/g)$}, or, including contributions from both constants $\alpha$ and $m_{\rm{e}}$ considered here: 
\begin{equation}
\frac{\delta f}{f}=(Q^{\rm{at}}_{\alpha}-Q^{\rm{L}}_\alpha)\frac{\delta \alpha}{\alpha}+(Q^{\rm{at}}_{m_e}-Q^{\rm{L}}_{m_e})\frac{\delta m_e}{m_e},
\label{eq:dff1}
\end{equation}%
\noindent where  $f=f_{\rm{L}}\approx f_{\rm{at}}$.  The frequency of the laser resonator is linear in the inverse resonator length: $f_{\rm{L}}\propto \,1/L \propto \, m_{\rm{e}}\alpha$ \cite{KozlovADP2019}. 
In addition, the  atomic frequency $f_{\rm{at}} \propto \, m_{\rm{e}}\alpha^{2+\epsilon}$, where the parameter $\epsilon$ accounts for enhanced sensitivity to $\alpha$ variation due to relativistic effects \cite{FlambaumCJP2009}. For the Cs D2 line employed in this work, $\epsilon\approx 0.26$ \cite{Dzuba2021}. Therefore, $Q_{\alpha}^{\rm{at}}=2.26$, $Q_{m_e}^{\rm{at}}=1$, $Q_{\alpha}^{\rm{L}}=1$ and $Q_{m_e}^{\rm{L}}=1$.

In applying Eq. (\ref{eq:dff1}), one has to distinguish  different frequency ranges that are determined by the various experimental time scales. The limit of low oscillation frequencies (probed, for example, in \cite{KennedyPRL2020}) is only one of the relevant ranges when probing rf oscillations \cite{KozlovADP2019,AntypasADP2020} as in this case additional ranges become relevant. For instance, the $f_{\rm{L}}$ follows changes in the resonator length up to the acoustic cut-off frequency of the resonator $f_{\rm{c1}}$, with $f_{\rm{c1}}\approx 50$\,kHz 
in our apparatus \cite{AntypasPRL2019}. At frequencies higher than $f_{\rm{c1}}$,  $f_{\rm{L}}$ is independent of the FC oscillations. This transition in sensitivity can be incorporated through a response function $h_L(f_{\rm \phi})$, with $h_L(f_{\rm \phi})$=1 below $f_{\rm{c1}}$ and $h_L(f_{\rm \phi})$=0  above $f_{\rm{c1}}$. In addition, the $f_{\rm{at}}$ is primarily sensitive to FC oscillations up to frequency $f_{c2}$ equal to the observed transition  
linewidth $\Gamma$. This atomic response can be characterized through the function $h_{at}(f_{\rm \phi})$, with $h_{at}(f_{\rm \phi})\rightarrow 1$ for $f_{\rm \phi} \ll f_{c2}$ and $h_{at}(f_{\rm \phi})\rightarrow 0$ for $f_{\rm \phi} \gg f_{c2}$. In practice, $h_{at}(f_{\rm \phi})$ is determined through apparatus calibration.  Inserting these response functions and the respective values of coefficients $Q_g$ into Eq. (\ref{eq:dff1}), one obtains:
\begin{equation}
\label{eq:dff2}
\frac{\delta f}{f}=\Big[2.26\,h_{\rm{at}}(f_{\rm \phi})-h_{\rm{L}}(f_{\rm \phi})\Big]\frac{\delta \alpha}{\alpha}+\Big[h_{\rm{at}}(f_{\rm \phi})-h_{\rm{L}}(f_{\rm \phi})\Big]\frac{\delta m_e}{m_e}.
\end{equation}

\noindent We see that $\delta f/f=1.26\,\delta \alpha/\alpha$ in the limit of low frequency $f_{\rm \phi}<f_{c1}$, \mbox{$\delta f/f=2.26\,\delta \alpha/\alpha +\delta m_e/m_e$} at intermediate frequencies $f_{c1}<f_{\rm \phi}<f_{c2}$, while \mbox{$\delta f/f\rightarrow 0$} in the limit of high frequency $f_{\rm \phi}\gg f_{c2}$.

If the FC oscillations arise due to  scalar UDM, their amplitude will be associated with couplings of the oscillatory UDM field to SM matter. 
This field is expected to exhibit stochastic amplitude fluctuations on time scale equal to its oscillation coherence time $\tau_{\rm c}$ \cite{Centers2021}. For measurement time $T\gg \tau_{\rm c}$ (such as in the present work  for the $\tau_{\rm c}$ values within the UDM models considered), this stochasticity can be neglected. The field  acquires a deterministic amplitude, and is given by\mbox{ $\phi(t)\approx{m_{\phi}}^{-1}\sqrt{2\rho_{\rm DM}}\sin({m_{\phi}t})$} \cite{BanerjeePRD2019},
%
%
where \mbox{$\rho_{\rm DM}\approx 3 \cdot 10^{-6}$}\,eV$^4$ is the estimated local galactic density of DM \cite{KimballBook}. Within the field, the constants acquire a small, time-dependent amplitude, such that:
\begin{equation}
\label{eq:alphaVar}
\alpha(t)=\alpha_0[1+g_{\gamma}\phi(t)],
\end{equation}
\begin{equation}
\label{eq:mVar}
m_e(t)=m_{e,0}\Big[1+\frac{g_e}{m_{e,0}}\phi(t)\Big],
\end{equation}
where and $g_{\gamma}$, $g_{e}$ are coupling constants of UDM to the photon and the electron,  and $a_0$, $m_{e,0}$ are the SM fine-structure constant and electron mass, respectively. One can make use of Eq.~(\ref{eq:dff2}) to relate an observed variation $\delta f/f$ to the couplings $g_{\gamma}$, $g_{\rm{e}}$:
\begin{equation}
\label{eq:deltafcases}
\frac{\delta f}{f}=\begin{dcases}
       1.26\,g_{\rm{\gamma}}{m_{\rm{\phi}}}^{-1}\sqrt{2\rho_{\rm DM}}h_{\rm{at}}(f_{\rm \phi}), & f_{\rm \phi}\leq f_{\rm{c1}} \\              \Big(2.26\,g_{\rm{\gamma}}+\frac{g_{\rm{e}}}{m_{\rm{e,0}}}\Big){m_{\rm{\phi}}}^{-1}\sqrt{2\rho_{\rm DM}}h_{\rm{at}}(f_{\rm \phi}), & f_{\rm \phi}>f_{c1}, 
    \end{dcases}
\end{equation}

\noindent where the atomic response $h_{\rm{at}}(f_{\rm \phi})$ is to be determined experimentally. In the low-frequency limit, there is no sensitivity to $g_{\rm{e}}$,  while above the acoustic cutoff $f_{c1}$ there is sensitivity to both $g_{\rm{e}}$ and $g_{\rm{\gamma}}$ couplings. In the absence of an observation of FC oscillations, Eq.\,\eqref{eq:deltafcases} can be used to place bounds on $g_{\rm{e}}$ and $g_{\rm{\gamma}}$, as it was done in \cite{AntypasPRL2019}.

\emph{Apparatus, data acquisition and analysis --} The experiment was designed to address the principal limiting factor of the previous work \cite{AntypasPRL2019} by introducing a more advanced data acquisition system. In addition, we implemented two different realizations of the setup in order to better control for spurious UDM signatures. The two setups  (Apparatus A and B) are described in the Supp. Mat.

Apparatus A is a new version of the
Cs Doppler-free polarization spectroscopy setup \cite{AntypasPRL2019}. The improvements include: a) using a stronger transition $6^{2} S_{1/2}(F= 4)\rightarrow 6^{2}P_{3/2}(F=5)$; b) increased laser-beam size and power to improve signal-to-shot-noise ratio; c) 
employing a graphics card to calculate and average card to efficiently process recorded data, 
in parallel with the data acquisition process.
The new apparatus features a nearly-100\% measurement duty cycle and can reach better statistical sensitivity in search for UDM
than that in Ref.\,\cite{AntypasPRL2019} in less than 1\,s (the experiment described in
\cite{AntypasPRL2019} took 66\,h in total).

Apparatus B was built independently of Apparatus A. It makes use of a different laser source, and implements Doppler-broadened spectroscopy of the $F=4\rightarrow F ^{\prime}=3,4,5$ components of the Cs D2 line, providing a bandwidth for the search for FC oscillations that is not limited by the transition natural linewidth ($\approx 5.5$\,MHz). Its data acquisition system samples the experimental signal at a lower rate compared to that in Apparatus A, resulting in a relatively lower sensitivity; however, this system is more immune to parasitic noise of technical nature.

In both experiments, sensitivity to FC oscillations is enabled by tuning the  laser in frequency to excite the respective atomic resonance. The spectroscopy signal is recorded in 1.1-s and 0.1-s long intervals for experiments A and B respectively, and corresponding power spectra are continuously computed and averaged. These are subsequently investigated for FC oscillations, that are expected to appear as excess power in the spectra. 

In Apparatus A 
we averaged 628700 power spectra corresponding to $\approx 187$\,h of pure acquisition time. Due to high resolution and statistical sensitivity, many thousands of spurious peaks are present in the resulting spectrum. The realization of Apparatus A does not allow us to eliminate them and we cannot establish a good UDM candidate exclusion at these points. Most of these peaks come from frequency modulation of the laser light, which is most probably the result of electromagnetic interference with a switching power supply (they form groups with peaks spaced by 20\,Hz). In this work, we present this apparatus as an ultimately sensitive device that requires some further design improvement in experimental technique as well as in data analysis 
\footnote{The raw-data file can be downloaded for analysis from the link  \url{https://irods-web.zdv.uni-mainz.de/irods-rest/rest/fileContents/zdv/project/m2_him_exp/mam/2022_dark_matter_v2/ExperimentA_data.dat?ticket=XD07Jrg8eBae4b7} 
(see technical description of the file in Supplementary Material).}.

In Apparatus B, we acquired data for a total of 113\,h, realizing a $\approx$16\% duty cycle. We alternated acquisition with the laser frequency tuned either on-, or off- the optical transition (where there is no sensitivity to FC oscillations), resulting in pure integration of 9 h in each case. This mixed data taking allows subtraction of the on- and off-resonance spectra and elimination of most of the signals
are not due to UDM. However,  a  total of 70 peaks remained in the subtracted spectrum with power exceeding a threshold for FC-oscillation detection at the 95\% confidence level (C.L.). These were primarily due to apparatus pickup
, or, due to parasitic laser frequency or amplitude noise. 
We investigated them using different methods [see Supp. Mat.]. For example, we did dedicated runs to check peaks that were nearly eliminated in the main run, and eventually observed residual power for these below the detection threshold. In addition, we took advantage of the in-tandem experiments to cross-check spurious UDM candidates. 
Several peaks in  the spectrum of `B', were either absent in `A', or had corresponding power significantly smaller than the detection threshold in `B', allowing elimination. Eventually, within the sensitivity of `B' we found no possible signatures of FC oscillations.

\begin{figure}
    \centering
    \includegraphics[width=\columnwidth]{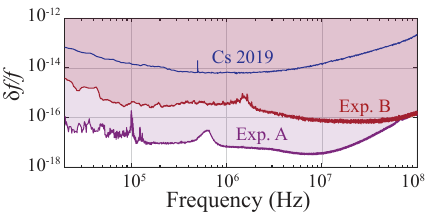}
    \caption{Constraints on the fractional frequency oscillations $\delta f/f$, shown for experiments A and B at the 95\% C.L, alongside  constraints from the earlier work \cite{AntypasPRL2019}. 
}    
    \label{fig:dfoverf}
\end{figure}

We show resulting $\delta f/f$ constraints in Fig.\,\ref{fig:dfoverf}, produced with consideration of the `look elsewhere' effect \cite{Scargle1982} for the $N\approx1.1\times10^8$ and $N\approx 1.6\times 10^6$ frequency bins in the power spectra of `A' and `B', respectively [see Supp. Mat.]. 
We note that the shown limits from `A' represent the ultimate apparatus sensitivity. This is likely achievable via a future implementation of a dual on- and off-resonance acquisition, and  cross-comparison of data with those from another, independent setup of similar sensitivity. %

\begin{figure*}
    \centering
    \includegraphics[width=\textwidth]{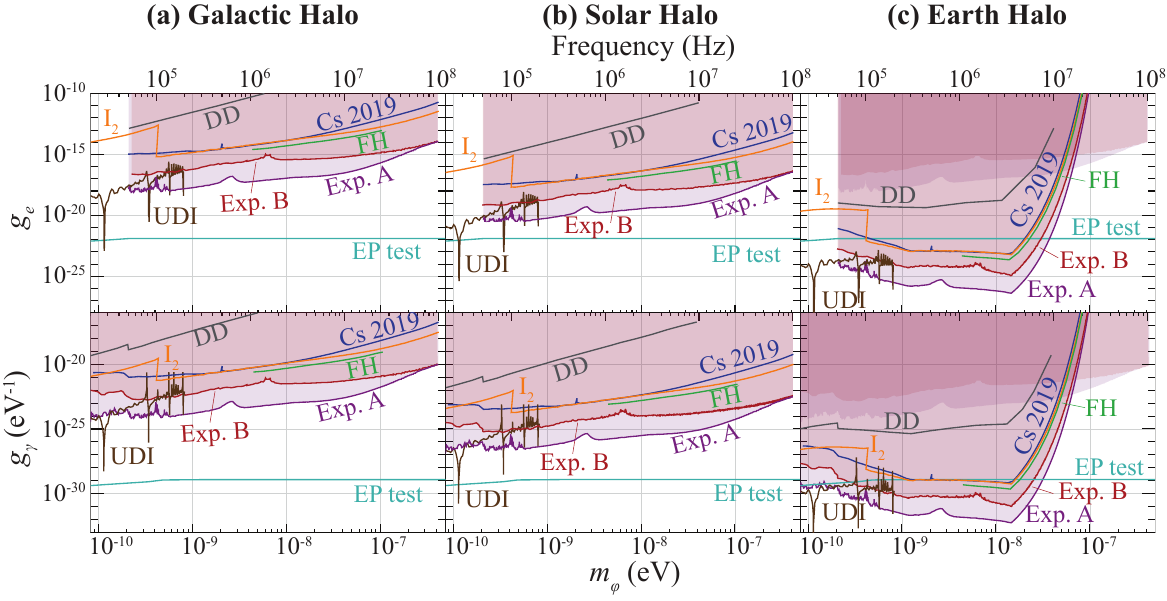}
    \caption{Exclusion plots at the 95\% C.L. for the coupling to the electron mass (top), and the fine-structure constant (bottom), produced within the Galactic-, Solar-, and Earth-halo UDM scenarios. 
    The constraint on $g_{\rm e}$ is only shown for $f>f_{c1}\approx50$\,kHz, [see Eq.\,(\ref{eq:deltafcases})]. Also shown are constraints from the previous work (Cs\, 2019) \cite{AntypasPRL2019}, Iodine spectroscopy~($I_2$)~\cite{OswaldArxiv2021}, an experiment using dynamic decoupling (DD) \cite{AharonyPRD2021}, an 
    unequal-delay interferometer (UDI) \cite{SavallePRL2020}, the Fermi lab Holometer (FH) \cite{aiello2021constraints}, and EP tests \cite{SmithPRD1999,SchlammingerPRL2008}. The exclusion regions for the Galactic-halo are also shown on the plot for the Earth-halo scenario, since these are stronger for the larger-mass region. 
  }
    \label{fig:final-plot}
\end{figure*}

\begin{figure}
    \centering
    \includegraphics[width=\columnwidth]{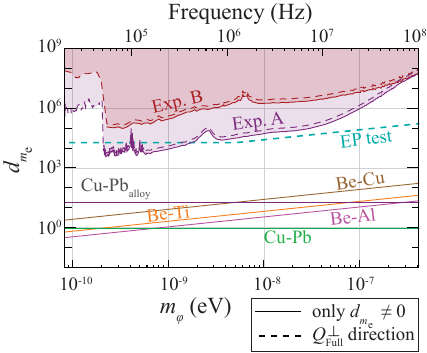}
    \caption{ Exclusion plot for $d_{m_e}$ at the 95\% C.L.; the solid lines assume a model where only $d_{m_e}\neq 0$. The dashed lines depict the bounds for a model defined by a vector of sensitivities, $\hat Q^\perp_{\rm Full}(m_\phi)\cdot \vec d$, that is orthogonal to the sensitivities of four leading EP test experiments projected onto $d_{m_e}$. 
    The bound from the fifth-best EP test experiment on the second model projected onto the $d_{m_e}$ direction, is shown by the dashed turquoise line. 
    } 
    \label{fig:dme_direction}
\end{figure}

\emph{Constraints on UDM couplings--} In the absence of detection of FC oscillations, we use the $\delta f/f$ constraints of experiment B (Fig.\,\ref{fig:dfoverf}), and apply Eq.\,(\ref{eq:deltafcases}) to set upper bounds on the UDM couplings to the electron mass $g_{\rm e}$ and fine-structure constant $g_{\rm \gamma}$, respectively. In addition, to illustrate the potential of our method in probing UDM, we consider constraints computed using the $\delta f/f$ limits  which may be ultimately feasible with experiment A (Fig.\,\ref{fig:dfoverf}).  

We show bounds for the case of the standard galactic UDM halo scenario ($\rho_{\rm DM}\approx 3 \cdot 10^{-6}$\,eV$^4$ \cite{KimballBook}), in Fig.\,\ref{fig:final-plot}\,(a). To derive these, we assume that FC oscillations arise due to a single coupling to either $g_{\rm e}$  or $g_{\rm \gamma}$, and incorporate a correction to account for degradation in sensitivity in the high-end of the investigated frequency range, due to the finite coherence of the UDM field (Q-factor of $\approx 1.1\cdot 10^6$ within the galactic halo scenario). 

The couplings $g_{\rm e}$ and $g_{\rm \gamma}$ can be further constrained within scenarios assuming the presence of a UDM halo that is gravitationally bound around the Sun \cite{ AndersonArxiv2020} or the Earth \cite{BanerjeeComPhys2020}. Within these scenarios, the  
UDM field has increased Q-factor, which is, respectively, $\approx 9\times 10^7$ and $\infty$. Relative to the standard galactic halo density, the UDM density is enhanced by $\approx \times 10^5$  for the Solar halo. For an Earth halo, the enhancement is strongly dependent on UDM-particle mass. 
 We show limits from consideration of these  models in Fig.\,\ref{fig:final-plot} (b) and (c).
 

An UDM field may couple to several species of the SM (this is indeed the case in the two concrete natural realization of scalar UDM, that were condsidered in the literature, either as a dilaton field~\cite{ArvanitakiPRD2015} or an axion subject to double breaking of the shift symmetry~\cite{FlackeJHEP2017,BanerjeePRD2019}). 
Thus, a UDM model can be described via a coupling-``vector" of five independent directions, $\vec d=d_{\alpha,m_e,\Lambda_{\rm QCD},(m_u+m_d)/2, m_d-m_u}$ in a five dimensional space, and a vector $\vec Q$ to quantify the respective sensitivity coefficients of any experiment.  
As noted in~\cite{OswaldArxiv2021}, the bounds arising from the direct UDM searches and EP tests are complementary to each other in this abstract space of coupling.
Consequently, one can 
find a direction $\hat Q^\perp_{\rm Full}(m_\phi)$
in the five dimensional parameter space that is orthogonal to the best four EP-test bounds for given mass. 
In our region of interest, $\hat Q^{\perp}_{\rm Full}(m_{\phi})$ is chosen as follows: below mass $5\times 10^{-9}\,$ eV it is orthogonal to the EP tests comparing two test bodies made out of  Be-Al~\cite{Wagner:2012ui}, Be-Ti~\cite{SchlammingerPRL2008}, Cu-Pb~\cite{SmithPRD1999}, Be-Cu~\cite{Su:1994gu} and written as $\hat Q^\perp_{\rm Full}(m_\phi)\simeq \big(
0.003\,,\, -0.987\,,\, 0.002\,,\,-0.001\,,\,-0.162\,\big)\,,$ and above $5\times 10^{-9}\,$ eV it is orthogonal to the Be-Al, Be-Ti, Cu-Pb, Cu-Pb alloy~\cite{Nelson:1990uk} EP tests and can be given as $\hat Q^\perp_{\rm Full}(m_\phi)\simeq \big(
0.020\,,\, 0.983\,,\, 0.018\,,\,-0.010\,,\,0.178\,\big)\,$. 
This choice of $\hat Q^{\perp}_{\rm Full}(m_{\phi})$ is the same as that discussed in \cite{OswaldArxiv2021}.
What is interesting is that $\hat Q^{\perp}_{\rm Full}(m_{\phi})$ has a sizable overlap with the direction of the electron coupling, $d_{m_e}$, which makes experiments looking for FC oscillations particularly powerful to search for this particular direction in coupling space.

Figure~\ref{fig:dme_direction} shows  bounds for the coupling to $m_{\rm e}$ along the direction $\hat Q^{\perp}_{\rm Full}(m_{\phi})$. (For consistency with \cite{OswaldArxiv2021}, we re-express  it in terms of the dimensionless constant  $d_{m_e} = g_e\Mpl/m_{e,0}$ where $\Mpl = \sqrt{\hbar c/({8\pi G_{\rm N}})}= 2.4\times 10^{18}$\,GeV is the Planck mass.) For comparison, we additionally show constraints from our direct UDM searches and EP tests, assuming only a UDM coupling to electrons. We see that in the direction  $\hat Q^{\perp}_{\rm Full}(m_{\phi})$, the direct search may ultimately approach (or surpass) the sensitivity level of EP tests.
In this direction, the present experiment A shows the potential of future direct UDM searches that can be used to probe  parameter space unconstrained by EP tests.

Focusing on a special direction in the multidimensional space that is orthogonal to the parameter space probed by EP tests represents a ``tuning'' of the model (or the direction of $\hat Q^{\perp}_{\rm Full}$) at the level of roughly 1:10$^3$, however, we still find it interesting as follows.
First, it highlights the value of pursuing different experimental approaches in parallel, as it is possible that our current theoretical biases are wrong and ``nature" chose this direction out of coincidence or just from other unknown theoretical reasoning (for analogous discussion see {\it e.g}~\cite{Arkani-Hamed:2006wnf,DiLuzio:2016sbl,Agashe:2006at,Fuchs:2020uoc,Arkani-Hamed:2021xlp,Balkin:2021rvh}, among many other works). 
Second, we would like to quantify the level of tuning and fine-tuning (a la 't Hooft~\cite{tHooft:1979rat}) required to define this model. 
Among the five-dimensional parameter space three, $d_{m_e,(m_u+m_d)/2, m_d-m_u}$, are technically natural and thus are radiatively stable, while $d_{\alpha,\Lambda_{\rm QCD}}$ are subjected to additive contributions. 
However, as mentioned above, in natural UDM models of the type of~\cite{ArvanitakiPRD2015,BanerjeePRD2019}, these additive contributions are under control at least to leading order by construction.
We can quantify the extra fine-tuning by looking at how much the presence of one coupling feeds into the other spoiling the delicate tuning. As the theory is perturbative, we can simply estimate as arising from one loop contribution for instance (omitting for simplicity logarithmic terms) 
$\Delta d_{\alpha}\sim {d_{m_e}\,\alpha/ 4\pi} ={\cal O} \left(10^{-3}\right) d_{m_e}\,,$ which implies only mild or no tuning. Similar conclusions apply to the strong sector upon replacing $d_{\alpha}$ with $d_{\Lambda_{\rm QCD}}$
as long as the scale that set the dark model coupling is larger than a few GeV. 
(Note that, if the scale is below GeV, the coupling to $\Lambda_{\rm QCD}$ does not receive any radiative correction.)

%


\emph{Conclusion --} The present results represent a sensitivity improvement in the direct search for ultralight scalar dark matter of up to three orders of magnitude with respect to earlier work.  

The sensitivity of the experiment is limited by our ability to suppress spurious noise; it might be possible to improve it by careful design of electronics and better electromagnetic shielding. Other future improvements may include designing an off-on resonance subtraction scheme in the higher-sensitivity Experiment A analogous to the one successfully implemented in Experiment B to suppress spurious spectral peaks. 
More importantly, both experiments together show that comparing two (or more) independent setups could be  an efficient way to suppress spurious peaks. Statistical sensitivity in these setups could be further increased by scaling up the vapor cell diameter. To obtain an optimal single-apparatus sensitivity in the whole investigated frequency range, one may employ both a narrow and a broad spectral line.
On the side of the theoretical interpretation, the existence of the special ``tuned'' directions in the parameter space where the present searches outperform EP tests, highlights the importance of pursuing different experimental approaches in parallel.   


The authors gratefully acknowledge A.\,Garcon, D.\,Kanta, and P.\,Otte for help with the project and V.\,V.\,Flambaum and V\,Dzuba for evaluating the relativistic corrections. 
AB thanks the Johannes Gutenberg University of Mainz for the hospitality and the Helmholtz Institute of Mainz for partial support during the completion of this work. 
This work was supported by the Cluster of Excellence ``Precision Physics, Fundamental Interactions, and Structure of Matter'' (PRISMA+ EXC 2118/1) funded by the German Research Foundation (DFG) within the German Excellence Strategy (Project ID 39083149), by the European Research Council (ERC) under the European Union Horizon 2020 research and innovation program (project Dark-OST, grant agreement No 695405) and Starting Grant, grant agreement No 947696), by the DFG Reinhart Koselleck project, and by Internal University Research Funding of Johannes Gutenberg-University Mainz.
The work of AB is supported by the Azrieli Foundation. 
The work of GP is supported by grants from BSF-NSF (No. 2019760), Friedrich
Wilhelm Bessel research award, GIF, the ISF (grant No.\,718/18), Minerva, SABRA-Yeda-Sela-WRC Program, the Estate of Emile Mimran, and The Maurice and Vivienne Wohl Endowment.

\bibliographystyle{apsrev4-1}
\bibliography{bibliography.bib}

\newpage

\newpage

\section*{Supplementary Information}

\subsection{Apparatus}
\label{"subsec:SI-apparatus"}
\begin{figure}
    \centering
    \includegraphics[width=\columnwidth]{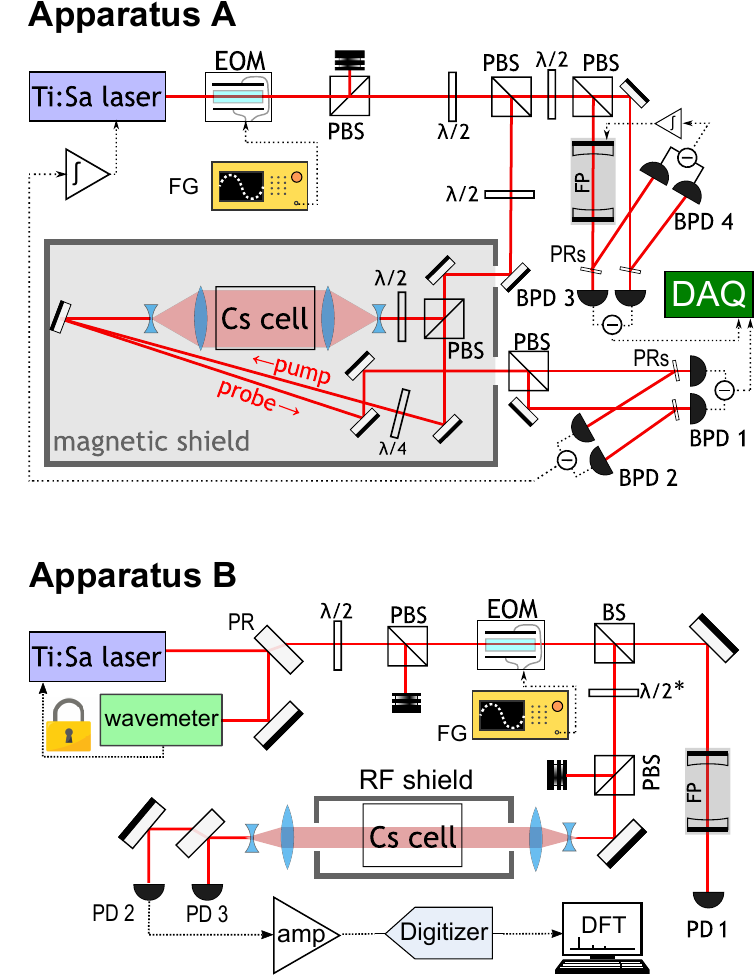}

  \caption{Experimental setups for the experiment A and B. Amp: amplifier; BPD: balanced photodetector; EOM: electro-optic modulator; DFT: discrete Fourier transform; FG: function generator; FP: Fabry-P\'erot optical cavity; $\lambda/2$: half-wave plate; PBS: polarizing beam splitter; PD: photodetector; PR: partial reflector; $\int$: integrator. 
The waveplate $\lambda/2^*$ is  motor-mounted and used to actively stabilize the optical power directed to the Cs cell. The data acquisition system (DAQ) of the Apparatus A is described in details in the Sec.\,\ref{subsec:"SI-DAQ"}.
} 
    \label{fig:experimental-setup}
\end{figure}

\emph{Experiment A --} The simplified schematic of the setup is shown in Fig.\,\ref{fig:experimental-setup}.

While based on the previously applied method \cite{AntypasPRL2019,AntypasADP2020}, the optical part of the experimental setup was fully revised. To increase the sensitivity of polarization spectroscopy, we increased the power of the pump and probe beams up to 5\,mW each. The photon shot noise at this power is at least two times higher than the total electronic noise of the photodetectors so the latter does not limit the measurements.
To avoid power broadening and keep the intensity similar to the previous experiment, the pump and probe beams diameters were increased up to 35\,mm. A correspondingly larger atomic vapor cell was used, with the inner diameter matching the beam sizes. We have also used a larger-size three-layer magnetic shield, with the entire polarization-spectroscopy setup built inside its innermost layer, with the exception of the optical polarization analyzer.

We used the strongest hyperfine transition of the $D_2$ line of Cs $6^{2}
S_{1/2}(F= 4) \longrightarrow 6^{2}P_{3/2}(F=5)$, which corresponds to 351.72196\,THz.

Fast (more than 100\,MHz bandwidth) photodetectors detecting the atomic-transition and FP etalon signals are connected to the digitizer directly with shortest possible low-losses double-shielded coaxial cables.

\begin{figure}
    \centering
    \includegraphics[width=\columnwidth]{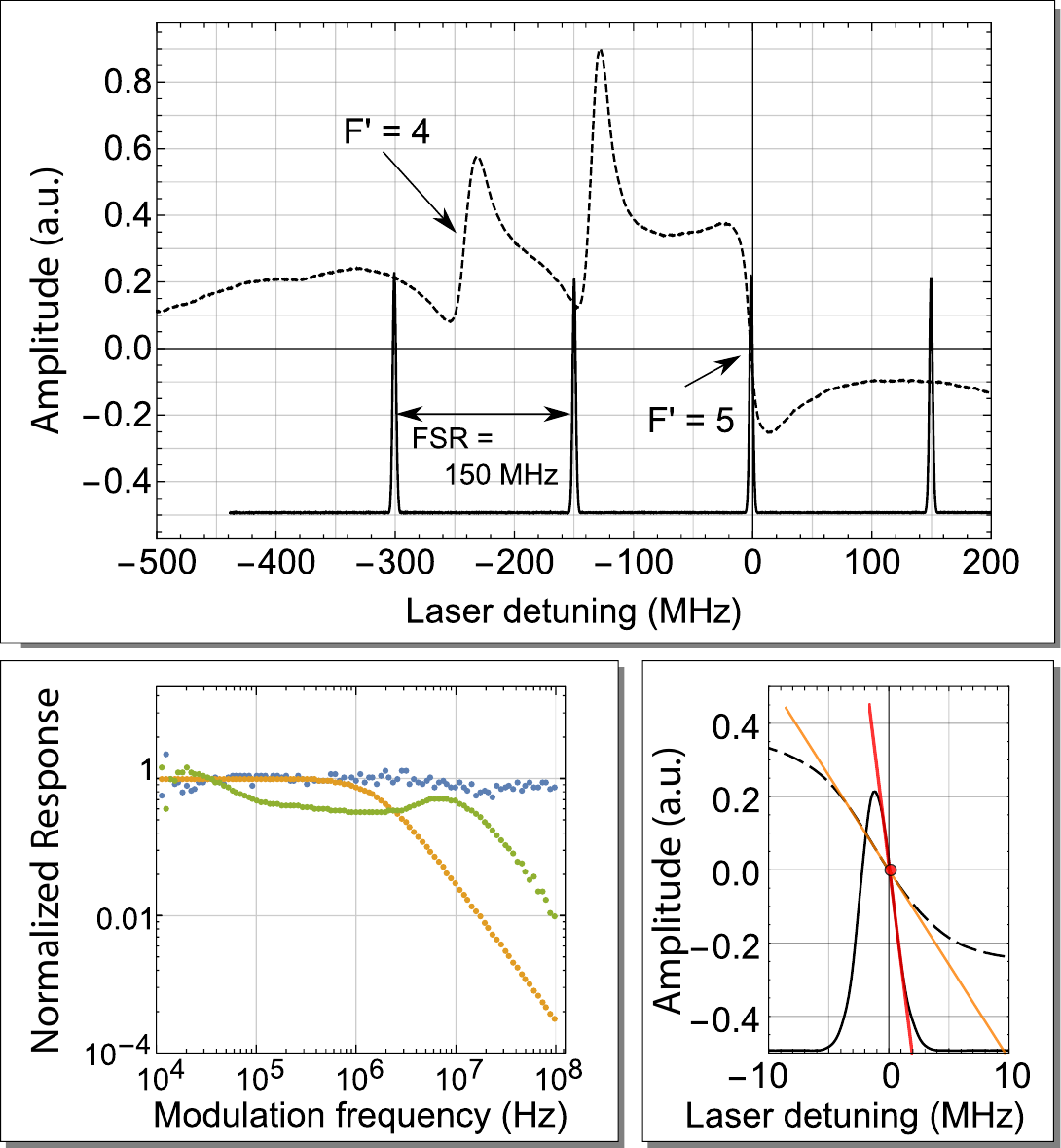}
    \caption{Top: dashed line -- polarisation spectrum of the $D_2$ transition, solid line -- FP transmission spectrum; bottom left: blue dots -- EOM transfer function, orange dots -- FP transfer function, green dots--atomic transfer function; bottom right: zoomed-in Cs $6^{2}S_{1/2}(F= 4) \longrightarrow 6^{2}P_{3/2}(F=5)$ transition (dashed line), FP resonance (solid line). Laser was locked to the atoms and FP cavity was locked to the laser at the point marked with the red dot. The modulation depth of the light frequency was estimated using the slope of the FP resonance line (marked with the red line). 
    }
    \label{fig:spectrum}
\end{figure}

A Ti:Sapphire laser was used as a light source, the same model as in the previous work. 
The laser was frequency-locked to the atomic polarisation spectroscopy setup. We used a separate balanced photodiode (BPD2 in Fig.\,\ref{fig:experimental-setup}) for the locking system to avoid the possibility of interference with measurement electronics (BPD1 in Fig\,\ref{fig:experimental-setup}). 
The reference FP cavity was locked to the laser using the side-of-fringe technique. The lock points for both systems are shown in Fig.\,\ref{fig:spectrum}.
Similar to the laser-locking system, we used independent photodetectors (BPD3 and BPD4 in Fig.\,\ref{fig:experimental-setup}).
The lock electronics were based on homemade low-noise analog integrators ($\int$) controlled with a microcontroller unit (MCU). We designed the feedback systems with a total bandwidth of less than 200\,Hz, such that possible dark-matter signatures in the range of several kHz and up addressed in this study are not affected.



\emph{Experiment B --} Apparatus B implements Doppler-broadened spectroscopy on the $F=4\rightarrow F ^{\prime}=3,4,5$ components of the Cs D2 line. Light from a T:Sapphire (not identical to that used in experiment A) laser excites atoms in a 25-cm-long Cs cell placed in an RF shield. The laser is similar to that used in experiment A, but has different amplitude and frequency noise spectrum. This allows for intercomparison of spurious signals in the two experiments, and elimination of such signals as UDM candidates. Sensitivity to UDM is enabled by tuning the laser frequency on the side of the atomic resonance, so that FC-oscillation-induced  $\delta f/f$ variations appear as amplitude variations on the light transmitted through the cell.  This transmission is additionally measured with the laser tuned off the Cs resonance, in order to record spurious apparatus signals present in the absence of sensitivity to UDM. 
 The transmission is recorded with a detector, whose output is recorded with a commercial digitizer in successive 0.1-s-long time series, every $\approx$\,0.5\,s.  
 After recording 400 time-series on resonance, we acquire the same number of series off resonance. This succession between on- and off-resonance is repeated continuously for a total of 113 hours of acquisition time (including data-transfer time and the time to tune the laser between the on- and off-resonance frequencies of $\approx 20$\,s). The net (on- plus off-resonant) data-acquisition time was  $\approx\, 18$ hours. The stored data are Fourier-analysed and the respective averaged power spectra with and without sensitivity to UDM are subtracted to obtain an `excess-power` spectrum which exhibits reduced number of apparatus spurious signals. This spectrum is analyzed for FC oscillations. 

\begin{figure}
    \centering
    \includegraphics[width=\columnwidth]{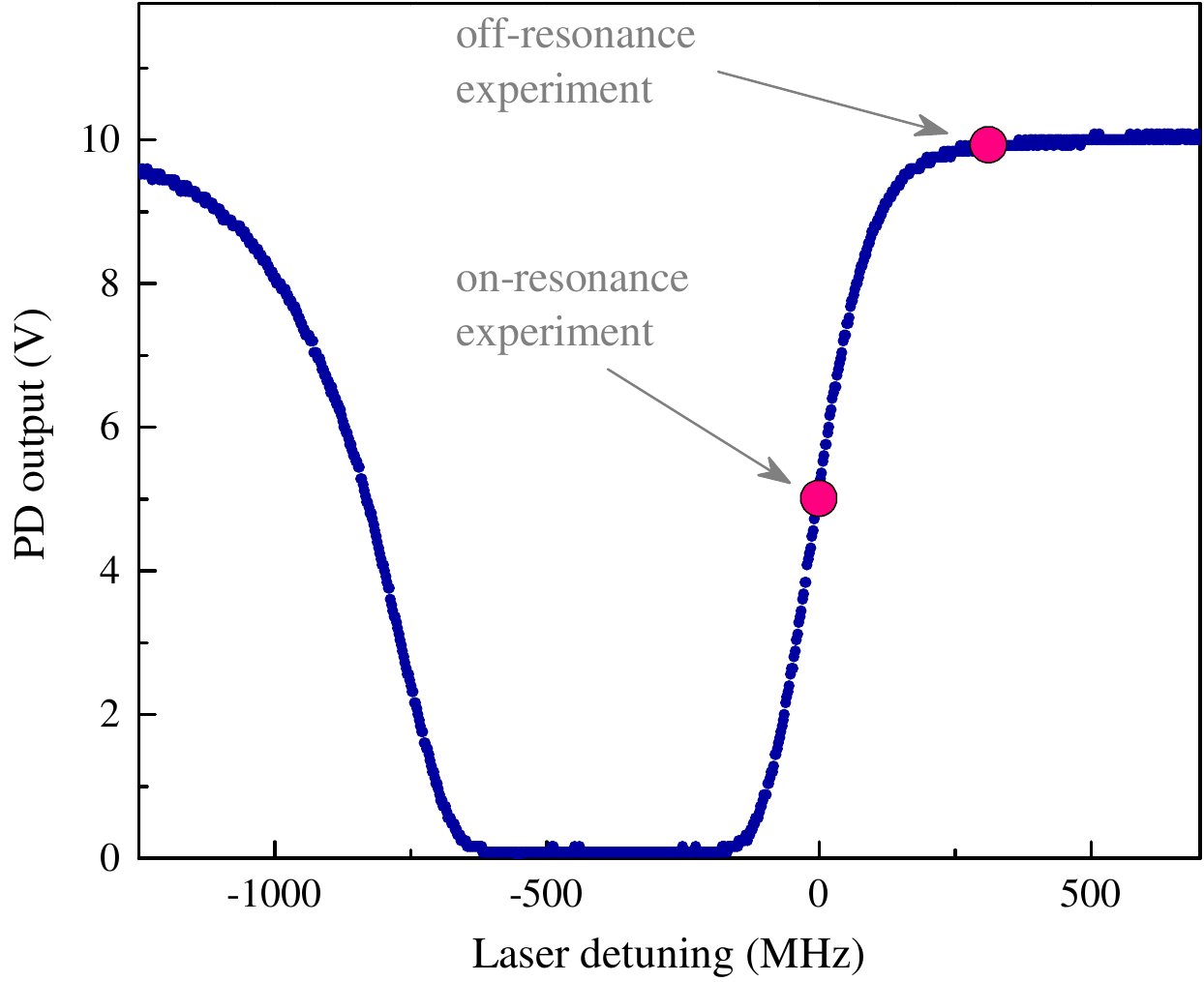}
  
    \caption{Spectrum of the Doppler-broadened $F=4\rightarrow F ^{\prime}=3,4,5$ transitions of the Cs D2 line, that are employed in experiment B. The magenta dots indicate the laser frequency lock points for data taken on or off the atomic resonance.}
    \label{fig:SpectrumB}
\end{figure}

The setup for experiment B  is shown in Fig.\,\ref{fig:experimental-setup}.  Light from the laser is directed to the Cs cell, and the part transmitted through is measured with a fast photodetector (PD2, model Thorlabs PDA10A). 
The light beam sent to the cell is $\approx$5\,mW in power and has a diameter of $\approx 12$\,mm. The cell is maintained at room temperature ($\approx 22^{\circ}$C) resulting in several absorption lengths at the resonance center (see Fig.\,\ref{fig:SpectrumB}). To probe  FC oscillations, the laser frequency is tuned to the side of the resonance and stabilized to the reading of a wavemeter (High Finesse WS8-2). For the data acquired off the Cs resonance, the laser frequency is detuned by 300\,MHz from the on-resonance value and feedback is applied to a motor-mounted half-wave plate so that  the power transmitted through the cell (as measured with the auxiliary photodetector PD3 shown in Fig.\,\ref{fig:experimental-setup}) is maintained to same level as that of the on-resonance transmission to within 0.1\%. This balancing of powers recorded on- and off- the Cs resonance aids in maintaining comparable spectral powers for the majority of spurious apparatus signals, so that in subsequent analysis for DM detection, most of these spurious signals can be removed from  the spectra. The fast detector output is amplified by $\times$100 with a commercial amplifier (Femto HVA-200M-40-B) and sent to the 12-bit digitizer described in Sec.\,\ref{subsec:"SI-DAQ"} 

A FP with resonance width (FWHM) of $\approx$150\,MHz and mirror spacing $\approx$10 mm is used to perform two tasks. First, it is employed to measure the amplitude of frequency modulation applied to the laser light with an EOM, as part of an atomic response calibration procedure that is described in Sec.\,\ref{subsec:"SI-DAQ"}. 
Second, the FP is used in auxiliary experiments to study frequency noise in the spectrum of the laser, as part of a process of characterizing spurious signals that are investigated for UDM detection (see Sec.\, \ref{subsec:SI-spurious-signal}) .

\subsection{Data acquisition system}
\label{subsec:"SI-DAQ"}

\begin{figure}
    \centering
    \includegraphics[width=0.5\textwidth]{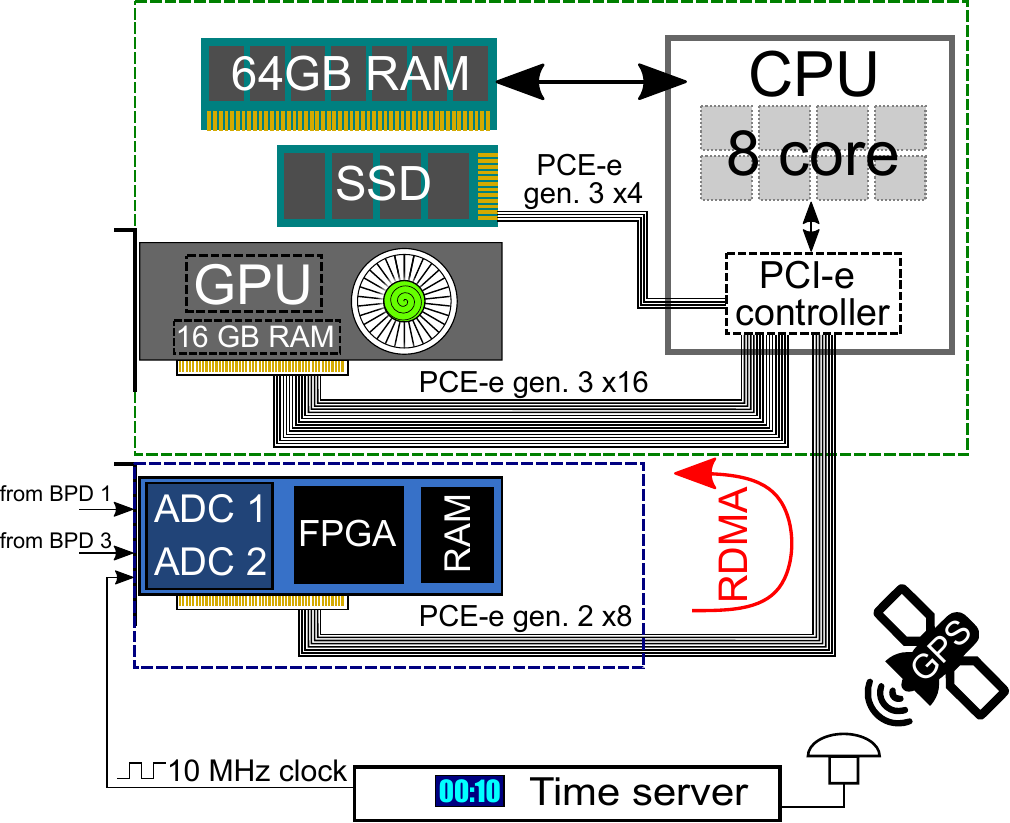}
    \caption{Data acquisition system (DAQ) diagram. The signals from balanced photodiodes BPD1 and BPD3 digitize by analog-to-digital converters ADC1 and ADC2 on the FPGA-based (field-programmable gate array) data acquisition card. The captured data uploads directly to the graphics card memory in RDMA (remote direct memory access) mode. The result of the Fourier transform on graphics processor unit (GPU) sends to host central processor unit (CPU) for the further averaging and saving on the solid-state drive (SSD). 
    }
    \label{fig:DAQ}
\end{figure}

\emph{Experiment A --} The previous experimental setup \cite{AntypasPRL2019} has been significantly updated to boost the sensitivity. The main difference is in the replacement of the heterodyne-type spectrum analyzer with one capable of operating in the frequency domain. The heterodyne technology sweeps the spectrum with a tunable oscillator whose frequency varies from the lower to the upper limit of the spectrum during a time interval on the order of milliseconds. Each (small) portion of the spectrum is then examined for a fraction of the sweeping period; the measurement is repeated only in the next scan. 
To collect the data in the previous experiment, the acquisition lasted over 60~hours but each frequency bin was analyzed only for a few milliseconds \cite{AntypasPRL2019,AntypasQST2021}.

For this reason, we followed a more modern approach, analyzing digitized sample sequences directly in the frequency domain. This approach is the same as implemented in the latest generation of commercial spectral analyzers. However, the sequence length of many commercial devices is limited up to $2^{16}=65536$ samples. Therefore, to overcome the limitations, we chose a two-channel FPGA-based digitizer (Spectrum M4i.4420-x8; FPGA stands for field-programmable gate array) with a sampling rate of 250\,MHz and a resolution of 16~bits. 
We use Meinberg LANTIME M600 timeserver as a 10\,MHz reference clock for analog-to-digital converters. This device based on oven-controlled crystal oscillator with GPS-synchronisation (GPS stands for global positioning system) has an accuracy averaged in 24\,h of better than $\pm 10^{-12}$.
With this setup we are able to analyze the spectrum up to 125\,MHz (the Nyquist frequency) with a resolution of 0.931\,Hz. The card is hosted in an eight-line 2-nd generation PCI-e (Peripheral Component Interconnect Express) slot of a standard personal computer that also runs the acquisition and analysis software. The digitizer driver supports the Remote Direct Memory-Access (RDMA) mode to transfer data into the the random access memory (RAM) of a graphics card (NVIDIA Quadro P5000) without the intervention of the CPU.

The data-taking sequence consists of the following steps:
\begin{enumerate}
    \item The digitizer sending data to graphics card RAM by the portions of 64MB of 16-bit integers data arrays;
    \item When the graphics card getting $2^{28}$ samples for each channel the 16-bit integers converting to float-point double precision numbers and running the fast Fourier transform (FFT) algorithm.  We used cuFFT library from NVIDIA CUDA Toolkit to perform FFT. To minimize memory usage we used the R2C (real-to-complex) FFT algorithm with writing the result on top of the initial data array;
	\item After the Fourier transform is done, we calculate the power spectrum in the same memory range on the graphics card and discard phase information;
	\item Then, we copy the resulting spectra from graphics' into the host's RAM and average it with previously measured spectra. We use OpenMP library to perform calculations on the CPU in the most parallel way and keep the CPU load homogeneous in time.
\end{enumerate}

We realised double buffering on graphics card such that when one buffer reaches the required amount of data ($2^{28}$ points) and Fourier transformation starts being performed on these data, we are filling the second buffer with the next portion of data. Thus, the data collection process takes place continuously without pauses for data processing.


A system based on commercial equipment and with a very favorable performance/cost ratio proved to be able to process online the acquired data, thus at a higher speed than acquisition, eliminating the dead time that plagued the previous experimental setup during the analysis. 
An analysis of the computational performance was conducted with the graphics card diagnostic tools (NVIDIA Visual Profiler) and showed a uniform use of the operating units and no overloading of the cores both graphics and CPU processors. 

The data acquisition system satisfied  our requirements of the bandwidth and data rate processing. It could be used for others applications where the real time spectrum analysis in a broadband (up to 250\,MHz) for two channels is required. 
Moving to higher bandwidth may require increasing the amount of memory on the graphics card to be able processing more than $2^{28}$ points. In case of using only one channel on the DAQ-card, or reducing of the data acquisition time, or switching from double (64-bit) to single precision (32-bit) floating point number calculations  we can double the bandwidth without upgrading the computer.

\emph{Experiment B --}The data acquisition system in experiment B is based on a commercial digitizer (PicoScope 5244D). This is used to acquire data in 0.1-s-long time series with 12 bit resolution at a rate of 250\,MSa/s. These data are stored in a computer and subsequently Fourier-analyzed. Approximately 0.4\,s is required to transfer each time series to the computer.

\subsection{Apparatus calibration}
\label{subsec:SI-calibration}

\emph{Experiment A --} Calibration of the frequency response was done using the electro-optical modulator (EOM in Fig.\,\ref{fig:experimental-setup}), as in Ref.\,\cite{AntypasPRL2019}
The EOM was aligned in a way to produce the light phase modulation without affecting polarization. We applied an AC voltage from the function generator (FG) to the EOM and minimized the signal on BPD1 at the modulation frequency. 
The calibration process was to apply AC voltage to the EOM with an amplitude inversely proportional to the modulation frequency. This means that the output light was frequency modulated with the same amplitude at modulation frequencies between 1\,kHz and 100\,MHz. 
The complete cycle of system calibration consisted of 124 measurements of the system response at frequencies uniformly distributed in the logarithmic scale.
We assumed that the FP cavity has a response of an ideal second-order filter \cite{tremblay1990frequency} and attributed the small deviations to imperfections of the EOM setup. Then we calibrated the EOM with respect to the FP. The modulation depth was estimated from the average ratio of amplitude modulation below the cut-off frequency divided by the FP resonance slope (red line in Fig.\,\ref{fig:spectrum}) to be $147\pm 4$\,Hz.
To avoid systematic error due to misalignment or laser-intensity drift we performed the calibration cycle at least every four hours. 
The  polarisation-spectroscopy response was described in \cite{torii2012laser}. In our experiment, 
it was found that the atomic transfer function could not be described as a perfect first- or second-order filter. It has a significant dip between 10\,kHz and 7\,MHz  dependent on the alignment and polarisation of the pump and probe beams. However, we performed the calibration based on the experimentally recorded atomic response 
linearly interpolated between neighboring points. 

\begin{figure}
    \centering
    \includegraphics[width=\columnwidth]{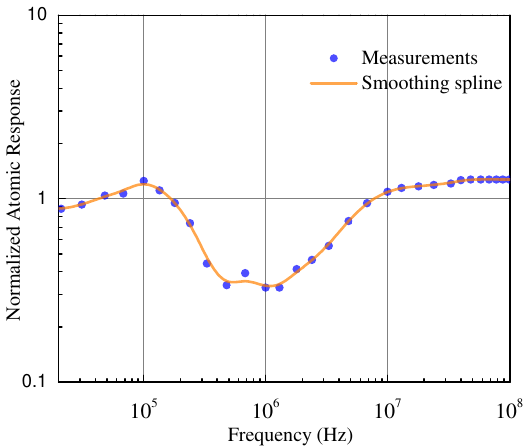}
  
    \caption{Normalized atomic response for Apparatus B. The smoothing spline applied to data points is used in subsequent analysis to compute the fractional frequency variations $\delta f/f$. 
    }
    \label{fig:AtomResponseB}
\end{figure}

\emph{Experiment B --} 
The response of Apparatus B to FC oscillations was calibrated with the method employed in A. Phase modulation was imposed on the laser light with an EOM (Fig. \ref{fig:experimental-setup}) and the resulting modulation experienced by Cs atoms was measured and compared to the modulation as measured with the FP. The latter was corrected (by as much as 9.5\% at 100 MHz) to account for the response attenuation of  the FP at high frequencies, as was also done in experiment A. The ratio of atomic to FP modulations was determined several times in the 20 kHz--100 MHz range employed in the present FC oscillation search. The average ratio represents the atomic response [function $h_{\rm at}$ of Eq. (\ref{eq:deltafcases})] and is shown in Fig.\,\ref{fig:AtomResponseB}. This atomic response is $\approx$1 in the limit of low and high frequencies, but is reduced by $\times$3 times around 1 MHz. Similar reduction was also observed in Apparatus A.  

\subsection{Data analysis}
\label{subsec:SI-data-analysis}
\emph{Experiment A --} 
During this experiment, we acquire data in strings of $2^{27}$ samples with a sampling rate of 250\,MSa/s. Fast Fourier transform was performed in real time with the graphics card (Sec.\,\ref{subsec:"SI-DAQ"}) in parallel with the data-acquisition process. No weighting function was applied to the data (square window). The frequency-bin size was 0.93\,Hz. Then, the resultant power spectra were averaged with the CPU and saved on the SSD drive. This setup allowed us to have a 100\% duty cycle of the experiment. We averaged 628700 power spectra in total, corresponding to $\approx 187$\,h of pure data-acquisition time. Every three-four hours or after any adjustment of the setup alignment, the calibration process (Sec.\,\ref{subsec:SI-calibration}) was repeated.

The spectrum recorded in Experiment~A can be downloaded from the link \url{https://irods-web.zdv.uni-mainz.de/irods-rest/rest/fileContents/zdv/project/m2_him_exp/mam/2022_dark_matter_v2/ExperimentA_data.dat?ticket=XD07Jrg8eBae4b7}. The file contains an array of $2^{27}$  double-precision floating point numbers (64-bit). Each number is the amplitude of the spectral component of the light intensity transmitted through the atomic medium normalized by the calibration function. It is given in Hz$_{RMS}$. The frequency bin is 0.931323\,Hz; the span is 0 to 125\,MHz. 

\begin{figure}
    \centering
    \includegraphics[width=\columnwidth]{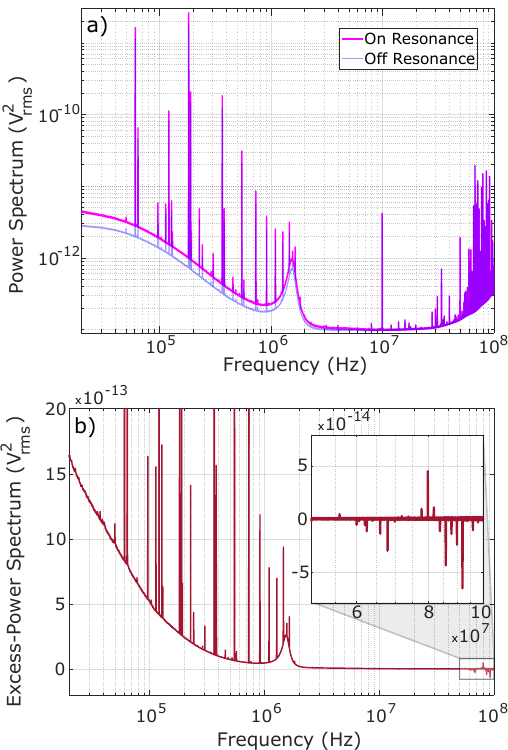}
  
    \caption{a) Averaged power spectrum shown for the on- and off-resonance acquisition done in experiment B.  
    b) Excess-power spectrum resulting from subtraction of the on- and off-resonance data of a). The spurious signals in the regions below and above 2\,MHz were investigated for DM detection in distinct ways (see text).}
    \label{fig:ExcessPowerSpectrum}
\end{figure}

\emph{Experiment B --} Each 0.1\,s time series obtained is split into four equally-sized 25\,ms segments and individually processed. Each segment is windowed in the time domain with a Hann window. Then, power spectra are obtained by applying a discrete Fourier transform to each segment and extracting the squared magnitude, with the resulting equivalent noise binwidth being 60\,Hz (due to windowing). All the power spectra for on- and off-resonance acquisition are averaged separately, and the results are shown in Fig.\,\ref{fig:ExcessPowerSpectrum}a. In order to eliminate spurious signals due to common-mode noise sources, we subtract the averaged on- and off-resonance power spectra and obtain the excess-power spectrum shown in Fig.\,\ref{fig:ExcessPowerSpectrum}b.

\subsection{Obtaining the 95\% confidence level threshold}

\begin{figure}
    \centering
    \includegraphics{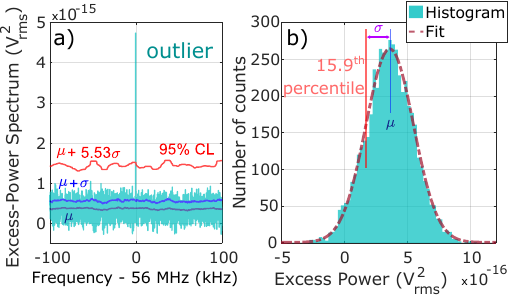}
    \caption{Example of outlier detection. a) Excess- power spectrum of experiment~B in a 200~kHz window centered at 56~MHz, near an outlier. The moving median $\mu$ and standard deviation $\sigma$ are calculated for a 12-kHz-sized window, as done in the analysis. b) The histogram (with a binwidth of $\approx 2.5\times 10^{-17}$~$\rm{V}^2_{\rm{rms}}$) of the excess power in the 200~kHz range shown in a) and a Gaussian fit. The standard deviation of the noise was obtained using the median and 15.9$^{\rm th}$ percentile as illustrated in b) and explained in the text. }
    \label{fig:ThreshExample}
\end{figure}

To search for candidate frequencies within the excess-power spectrum, we compared the power in each frequency bin to the noise in its vicinity, as depicted in Fig.\,\ref{fig:ThreshExample}. The resulting noise in the averaged spectra is well described by a Gaussian (shown in Fig.\,\ref{fig:ThreshExample}b), as expected from the central limit theorem. The standard deviation of this noise gives a natural scale to set a threshold to discriminate outliers. Additionally, we used this threshold to put limits on the excess power variations for experiment~B. This threshold was also calculated and presented for experiment~A as the sensitivity region, these are not limits, as there is a large amount of peaks of unknown origin, 
see, for example, Fig.\,\ref{fig:LowFreqSpurious}.

Before obtaining the moving standard deviation of the excess power spectrum $\sigma$, we removed the baseline of the spectrum, as the slope of the baseline can lead to artificially overestimating $\sigma$. The procedure to obtain the baseline was the same for experiments~A and B and consisted of 1)~applying a median filter to eliminate outliers and 2)~a second-order Savitzky-Golay filter for further smoothing. We used filter-window sizes 3\,kHz and 12\,kHz for experiments~A and B, respectively. We confirmed that this baseline-removing procedure leaves $\sigma$ unchanged to within $>$0.2~\%, by applying it to synthetic Gaussian noise. However, this procedure could not be used for experiment~A at low frequencies ($<$200~kHz) because of the high density of large peaks, so this region was treated differently: outliers and problematic ranges were manually removed and then only the Savitzky-Golay filter was applied to produce the smooth baseline.

After the baseline is subtracted, $\sigma$ is extracted using rank-order filters to obtain the moving median, $\mu$, and the 15.9$^{\rm th}$ percentile. The subtraction of the median and the 15.9$^{\rm th}$ percentile yields an estimator for the local $\sigma$. We preferred this to methods using the variance and mean, as these quantities are more susceptible to large outliers. Having characterized $\sigma$, it is possible set the 95\% confidence level threshold, $P_{\rm{th}}$, so that if a spectral feature has excess power $P_{\rm{ex}}>P_{\rm{th}}$, there is probability $p_0=5\%$ that it is due to statistical fluctuations.

Naively, one would set the threshold at around 2$\sigma$. However, this has to be adjusted for the fact there are $N$ bins in the spectrum, and a relatively large fraction ($\approx 5\%$) of them is expected to have power in excess of $P_{\rm{th}}$ \cite{Scargle1982}. Accounting for this effect, the threshold will be given by:
\begin{equation}
    \label{eq:Pth}
    P_{\rm{th}} = \sqrt{2}\rm{erf}^{-1}\Big\{2(1-p_0)^{1/N}-1\Big\}\sigma + \mu,
\end{equation}
so $P_{\rm{th}} - \mu$ is 6.15$\sigma$ for experiment~A and 5.53$\sigma$ for experiment~B. From this threshold $P_{\rm{th}}$, the frequency variation $\delta f$ can be constrained. The variation is given by $\delta f=\delta V/D$, where $\delta V$ is the voltage variation corresponding to the variation $\delta P$ in the excess power spectrum of Fig.\,\ref{fig:ExcessPowerSpectrum}b (at the 95\% confidence level). The average discriminator slope was $D\approx$ 39.8\,mV/MHz for experiment~B. The slope was consistent within 4\% throughout the data taking.
The quantities  $\delta V$ and  $\delta P$ are related via $\delta P=2V\delta V=2\sqrt{P_{\rm{av}}}\delta V$, where $P_{\rm{av}}$ is the averaged power spectrum recorded with on-resonance acquisition.
This results in a limit for $\delta f$ given by
\begin{equation}
\label{eq:deltafSupplMat}
\delta f=\frac{\delta V}{D}=\frac{\delta P}{2\sqrt{P_{\rm{av}}}D}.
\end{equation}

\emph{Scalloping losses --} In the regime where the expected FC oscillations are coherent for periods of time much longer than the integration time, the linewidth of the FC oscillations are much narrower than the frequency binwidth in the acquired power spectra. If the frequency of these oscillations lies between two of the frequency bins in the spectra, the power due to these oscillations will be spread among the bins, leading to scalloping losses~\cite{Sevgi2007,Heinzel2002} and a corresponding decrease in sensitivity to FC oscillations. In order to account for this effect, we have assumed a constant sensitivity loss given by the average normalized power response that would be observed for an oscillation with narrow linewidth, which corresponds to $\approx$77\% and $\approx$90\% for experiments A and B, respectively. The difference between the experiments comes from the use of the different windows: rectangular and Hann for experiments A and B, respectively.

\emph{Decoherence losses --}
Another loss mechanism is related to the opposite regime---when the FC oscillations have a linewidth that is larger than that of the frequency binwidths of the acquired power spectra. In this case, the power due to the FC oscillations is spread over more than one frequency bin. This leads to worse limits in the case when one looks for excess power in individual bins, because a single bin captures a smaller fraction of the total FC-oscillation power. 
The sensitivity degradation is only relevant for the scenario involving the standard galactic halo model~($Q\sim 10^6$), as for the Solar~($Q\sim10^8$) and Earth halo~~($Q\gg10^8$) the expected FC-oscillation linewidths are narrower than the binwidths in our acquired power spectra over the entire frequency range for both experiments.

In order to quantify the loss in the standard galactic halo scenario, we considered the expected lineshape~$f_{DM}$ from the FC oscillations in this case. The width of~$f_{DM}$ arises from frequency shifts due to the virialized motion (with a velocity dispersion of $v_0 \approx 10^{-3} c_{\rm{0}}$, where $c_0$ is the speed of light in vacuum of the UDM particles (as derived in Refs.~\cite{Turner1990,Foster2018, Gramolin2021}),

\begin{equation}
f_{\rm UDM}(f) = \frac{2c_{\rm{0}}^2}{\sqrt{\pi} v_0 v_{\text{lab}} f_\phi} \exp{\left(-\frac{2c_{\rm{0}}^2}{v_0^2} \frac{f - f_\phi}{f_\phi} - \frac{v_{\text{lab}}^2}{v_0^2}\right)} \sinh{\beta}, \label{eq:scalar_lineshape}
\end{equation}
where $v_{\text{lab}}$ is the velocity of the laboratory in the galactic frame (taken to be $\approx 230$\,km/s), and we have denoted

\begin{equation}
\beta =\frac{2c_{\rm{0}} v_{\text{lab}}}{v_0^2} \sqrt{\frac{2(f - f_\phi)}{f_\phi}}
\end{equation}
for brevity, and $f_\phi$ is the Compton frequency of the DM particle. 

The fraction of the power captured by a frequency bin can then be estimated by integrating the lineshape over the binwidth (we approximated the bin response to that of a rectangle with the same width). This quantity is dependent on the bin's center frequency, but the maximum possible fraction (ie.~when the bin is optimally placed) can be easily obtained by convolving the lineshape with the binwidth and finding the maximum. However, it is unlikely that the frequency bins in the acquired power spectra align with the optimal placement; in order to account for this we additionally multiplied this sensitivity loss ratio with the scalloping loss factors discussed above. This final sensitivity loss factor is then calculated for different UDM Compton frequencies and used to soften the limits. This is a conservative estimation as the losses due to the suboptimal centering of the bins is expected to be smaller that the scalloping losses. This is because the FC-oscillation-linewidths are broader that the frequency binwidths, so the losses between optimal and least optimal relative positions of the FC-oscillation frequency and bin-center frequency are smaller than the scalloping losses that would be expected in the coherent scenario.




\emph{UDM field amplitude stochasticity--} As mentioned in the main paper, one has to  generally consider the stochastic fluctuations of the UDM amplitude \cite{Centers2021}. This effect is important in analysis in cases where the experimental measurement time $T$ is smaller than the field's coherent oscillation time $\tau_{\rm c}$. In the experiments A and B, we have $T=$187 h and 114 h, respectively. For the standard- galactic and Solar-halo scenarios (with $Q\approx 1.1\times 10^6$ and $\approx 10^8$, respectively), the longest relevant coherence time (corresponding to the lowest measured frequency of 20 kHz), is  $\approx 9$ s and $\approx 800$ s, respectively. Since $T\gg \tau_{\rm c}$ throughout the investigated frequency range, the stochasticity can be neglected and one may use the deterministic value of the UDM field amplitude $\phi_{0}=m_{\phi}^{-1}\sqrt{2\rho_{\rm DM}}$. 
In the Earth halo model, the coherence time is given by  $\tau_{c}=1/ m_{\phi}\upsilon^2$ \cite{BanerjeeComPhys2020}, where $\upsilon\approx 10^4$ m/s is the UDM particle velocity. For the lowest frequency of 20 kHz  ($m_{\phi}=8\times 10^{-11}$eV), we obtain $\tau_{\rm c}\approx$1.5\,h. Therefore, within the Earth halo model as well, the condition $T\gg \tau_{\rm c}$ is satisfied and stochasticity may be neglected. 

\subsection{Spurious signals}
\label{subsec:SI-spurious-signal}
As discussed in the main text, experiments A and B are complementary. While Apparatus A has higher sensitivity to FC oscillation detection, the acquisition modality employed in B, involving both on- and off-resonance measurements, results in reduced number of spurious signals to be investigated for UDM detection. This investigation was done in two distinct frequency regions: a region below, and another above 2\,MHz. 

\begin{figure}
    \centering
    \includegraphics[width=\columnwidth]{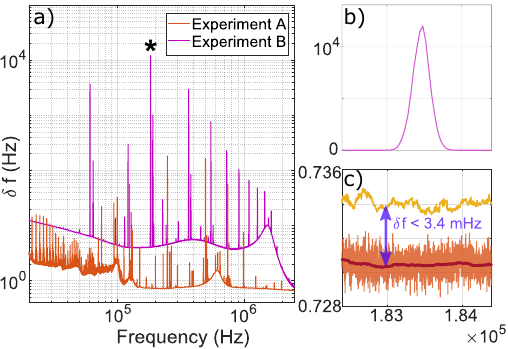}
    \caption{Example of the candidate-elimination strategy used below 2\,MHz. The spectra of both experiments are shown in a). The largest outlier in experiment~B is near 183\,kHz marked with an asterisk). b) and c) are zoomed portions of the spectra for experiments~B and A, respectively, around the center frequency of the spurious peak. The absence of a corresponding peak in experiment~A allows us to place DM limits at this frequency.}
     \label{fig:LowFreqSpurious}
\end{figure}

In the region below 2\,MHz, the spurious signals in experiments A and B were intercompared. Within this region the data of setup B contain 42 peaks with excess power above the 95\% detection threshold for DM. It was found that all but one of these peaks are not present in the spectrum of A (see Fig.\,\ref{fig:LowFreqSpurious}). Therefore they could be readily eliminated from being UDM candidates. A single spurious signal at frequency of 97040 Hz was observed in both data sets. This was found to be due to laser frequency noise. Its power was measured in an auxiliary experiment with Apparatus B using both the atomic and the FP resonance. When the respective powers were subtracted, the residual excess power was below  the  95\% detection threshold; thus the peak was eliminated from being a UDM candidate. 

In the region above 2\,MHz, the excess-power spectrum of experiment B (Fig.\,\ref{fig:ExcessPowerSpectrum}) contained a total of 28 spurious peaks. These were checked one by one in auxiliary experiments with the laser tuned on or off the atomic resonance. A commercial spectrum analyzer (Keysight N9320B) was used to acquire power spectra in this work, as it is more efficient that the digitizer employed in the main experiment, when analyzing small frequency windows. The excess power of all spurious peaks after subtraction of the `on' and `off' powers was below the set UDM detection threshold. Thus, they were eliminated from being UDM candidates. Their origin was attributed to apparatus pickup, such as for example, intrinsic noise of the digitizer or environmental rf noise. Note that several of the spurious peaks in the excess-power spectrum (see inset of Fig. \ref{fig:ExcessPowerSpectrum}) have negative power. This is because of the stochastic behavior of the difference of powers in the `on-' and `off-' resonance acquisitions. Naturally, the excess negative power cannot arise due to FC oscillations. These peaks, however, were also investigated with auxiliary experiments in order to eliminate the presence of FC oscillations at the reported levels. 


\end{document}